\documentclass[aps,prb,twocolumn,superscriptaddress,showpacs]{revtex4-1}
\usepackage{amsmath}
\usepackage{color}
\usepackage{amssymb}
\usepackage{epsfig}
\usepackage{multirow}
\usepackage{amsbsy}
\usepackage{array}
\usepackage{diagbox}
\usepackage{bm}
\usepackage{extarrows}
\usepackage{graphicx}
\usepackage{appendix}
\usepackage[normalem]{ulem}
\usepackage{txfonts}

\usepackage[colorlinks=true,linkcolor=blue,citecolor=blue,urlcolor=blue,bookmarks=false]{hyperref}

\begin{document}
	
\title{Efficient tensor network representation for Gutzwiller projected states of paired fermions}

\author{Hui-Ke Jin}
\affiliation{Department of Physics, Zhejiang University, Hangzhou 310027, China}
\affiliation{Kavli Institute for Theoretical Sciences $\&$ CAS Center for Excellence in Topological Quantum Computation, University of Chinese Academy of Sciences, Beijing 100190, China}

\author{Hong-Hao Tu}
\email{hong-hao.tu@tu-dresden.de}
\affiliation{Institut f\"ur Theoretische Physik, Technische Universit\"at Dresden, 01062 Dresden, Germany}

\author{Yi Zhou}
\email{yizhou@iphy.ac.cn}
\affiliation {Beijing National Laboratory for Condensed Matter Physics $\&$ Institute of Physics, Chinese Academy of Sciences, Beijing 100190, China}
\affiliation{Songshan Lake Materials Laboratory, Dongguan, Guangdong 523808, China}
\affiliation{Kavli Institute for Theoretical Sciences $\&$ CAS Center for Excellence in Topological Quantum Computation, University of Chinese Academy of Sciences, Beijing 100190, China}

\date{\today}
\begin{abstract}
Recent work by Wu {\em et al.} [arXiv:1910.11011] proposed a numerical method, so-called matrix product
operator-matrix product state (MPO-MPS) method, by which several types of quantum many-body wave functions, in particular, the projected Fermi sea state, can be efficiently represented as a tensor network.
In this paper, we generalize the MPO-MPS method to study Gutzwiller projected paired states of fermions, where the maximally localized Wannier orbitals for Bogoliubov quasiparticles/quasiholes have been adapted to improve the computational performance.
The study of $SO(3)$-symmetric spin-1 chains reveals that this new method has better performance than variational Monte Carlo for gapped states and similar performance for gapless states.
Moreover, we demonstrate that dynamic correlation functions can be easily evaluated by this method cooperating with other MPS-based accurate approaches, such as the Chebyshev MPS method.
\end{abstract}

\maketitle

\section{Introduction}

In physics, a large family of states of matter can be described by paired fermions, which range from superconductors~\cite{bcs} and superfluids~\cite{Leggett75} to nuclei and neutron stars~\cite{neutronstar}. The ground states and low-energy excited states of these paired fermions are governed by an effective Hamiltonian of Bogoliubov-de Gennes (BdG) type, and corresponding ground-state wave functions are of Bardeen-Cooper-Schrieffer (BCS) type. The idea of paired fermions was also combined with Gutzwiller projection by Anderson to describe doped Mott insulators and quantum spin liquids, which is dubbed as ``resonating valence bond" (RVB)~\cite{anderson73,Anderson1987,anderson04,rmp06,QSLRMP}. In practice, the Gutzwiller projection is performed numerically by the variational Monte Carlo (VMC) method and the physical observables can be evaluated accordingly~\cite{Gros89}. As a result, these Gutzwiller projected wave functions serve as a rather good variational ansatz for strongly correlated electrons and quantum spin systems~\cite{QSLRMP}.

Meanwhile, tensor networks provide an alternative way to construct variational wave functions for quantum many-body systems. Some of the most popular tensor networks, such as the matrix product state (MPS), the projected entangled-pair state (PEPS), and the multiscale entanglement renormalization ansatz, are widely used as the basis for variational approaches to quantum many-body problems~\cite{VCM08,CiracVerstraeteJPA09,schollwock11,vidal07mera,Orus13}. Considering the success of both methods (Gutzwiller projected states and tensor network states), it is very natural for us to ask the question: Could we establish some generic relations between these two methods? However, the general relation between Gutzwiller projected states and tensor network states remains unclear so far, despite the fact that some RVB states have exact PEPS representations~\cite{Verstraete06a,Schuch12,Wang13,Poiblanc13,yang12,wildeboer12}.

Very recently, it was realized by Wu {\em et al.}~\cite{wu2019} that Gutzwiller projected Fermi sea states can be expressed as tensor network states by rewriting a linear combination of single-particle operators as a matrix product operator (MPO) with bond dimension $D=2$. In this way, a Gutzwiller projected Fermi sea state can be obtained by applying the MPOs to an MPS one by one, which is termed as ``MPO-MPS method''. Thus, the Gutzwiller projected Fermi sea state can be efficiently expressed as a tensor network state. The efficiency of this tensor network representation has been examined carefully with several paradigmatic wave functions in Ref.~[\onlinecite{wu2019}]. If one chooses the basis of the single-particle state properly, namely, by using {\em  maximally localized Wannier orbitals}, then the performance of the MPO-MPS method will be improved dramatically. The immediate advantage of the MPO-MPS method is two-fold: (i) For Gutzwiller projected states, the computation of various important characteristic quantities, such as entanglement spectrum and von Neumann entanglement entropy, becomes possible under the MPS representation. (ii) For MPS-based variational approaches, such as density matrix renormalization group (DMRG), the Gutzwiller projected states could be used as a good initial input to speedup numerical simulations.

In this paper, we shall generalize the basis-optimized MPO-MPS method to study Gutzwiller projected paired states of fermions in one dimension. Indeed, we will demonstrate that the basis-optimized MPO-MPS method is equally efficient to compute the Gutzwiller projected paired states as well as the unprojected paired states, since the Gutzwiller projection can be viewed as an MPO with bond dimension $D=1$. The method developed here complements the basis-optimized MPO-MPS method proposed in Ref.~[\onlinecite{wu2019}] for the Gutzwiller projected Fermi sea state. This completes the basis-optimized MPO-MPS toolbox for Gutzwiller projected fermionic wave functions. Furthermore, we demonstrate that the MPO-MPS method can be used to calculate dynamic correlation functions with the help of other MPS-based accurate approaches, such as the Chebyshev MPS method~\cite{chebyshevmps}.

The remaining part of this paper is organized as follows.
In Sec.~\ref{sec:BdG}, we introduce a generic BdG Hamiltonian and diagonalize it by a generalized Bogoliubov transformation. The ground state of paired fermions is obtained by filling all the Bogoliubov quasiholes, on which the Gutzwiller projection will be implemented.
In Sec.~\ref{sec:mpo-mps}, the MPO-MPS method is formulated for paired fermions and the implementation of the Gutzwiller projection is discussed.
In Sec.~\ref{sec:TFXY}, we use one dimensional (1D) transverse field XY model as a benchmark to compare various MPO-MPS methods.
In Sec.~\ref{sec:bbq}, $SO(3)$-symmetric $S=1$ spin chains are studied in details. The truncation error of the MPO-MPS method is estimated. The ground-state energy, the von Neumann entanglement entropy and the spin spectral function are computed.
Section \ref{sec:summary} is devoted to summary and discussions.

\section{BdG Hamiltonian and the ground state of paired fermions}\label{sec:BdG}

We start with a generic BdG Hamiltonian that describes paired fermions as follows,
\begin{equation}
H_{\text{BdG}} = \sum_{k,l=1}^{N}t_{kl}c_k^{\dagger}c_{l}+\frac{1}{2}\sum_{k,l=1}^{N}(\Delta_{kl}c_kc_{l} + \mathrm{h.c.}),
\label{eq:BdG}
\end{equation}
where $k,l=1,\ldots,N$ denote generic single-particle degrees of freedom, such as lattice sites and spin/flavor indices, and $c_{k}$ and $c_{k}^{\dagger}$ are fermion annihilation and creation operators, respectively.
Note that $t_{kl}=t^{*}_{lk}$ ($*$ denotes complex conjugate) due to the Hermiticity of the Hamiltonian and $\Delta_{kl}=-\Delta_{lk}$ because of the fermion anticommutation relation.

For short, we define the vector operator $c=(c_{1},\ldots,c_{N})^{T}$ and the matrices $t$ (with elements $t_{kl}$), and $\Delta$ (with elements $\Delta_{kl}$) and introduce the Nambu representation $\psi^{T}=(c^{T},c^{\dagger})$. In the Nambu representation, the Hamiltonian $H_{\text{BdG}}$ can be written in the matrix form
\begin{equation}\label{eq:BdGM}
H_{\text{BdG}} = \frac{1}{2}\psi^{\dagger}\mathcal{H}_{\text{BdG}}\psi + \frac{1}{2}\mathrm{tr}(t)
\end{equation}
with
\begin{equation}
\mathcal{H}_{\text{BdG}}=\left(\begin{array}{cc} t & \Delta \\ -\Delta^{*} & -t^{*}\end{array}\right).
\end{equation}
The matrix $\mathcal{H}_{\text{BdG}}$ can be diagonalized by a $2N\times 2N$ unitary matrix $\mathcal{M}$ as follows:
\begin{equation}
\mathcal{M}^{\dagger}\mathcal{H}_{\text{BdG}}\mathcal{M} = \left(\begin{array}{cc} -\Lambda & 0 \\ 0 & \Lambda\end{array}\right),
\end{equation}
where $\Lambda$ is a diagonal and non-negative matrix characterized by matrix elements $\Lambda_{kl}=\varepsilon_{k}\delta_{kl}$, and $\mathcal{M}$ satisfies the unitary relation $\mathcal{M}^{\dagger}\mathcal{M}=\mathcal{M}\mathcal{M}^{\dagger}=1_{2N}$ and is of the form
\begin{equation}
\mathcal{M}=\left(\begin{array}{cc}
V & U^{*}\\
U & V^{*}
\end{array}\right),
\end{equation}
where $U$ and $V$ are two $N\times{}N$ matrices satisfying the relations $V^{\dagger}V+U^{\dagger}U=1_{N}$ and $U^{T}V+V^{T}U=0$. Here $\varepsilon_{k}\ge 0$ is the Bogoliubov quasiparticle excitation energy. It is worth noting that we have chosen the quasihole representation for later convenience so that the Bogoliubov quasihole creation operators $d_{m}^{\dagger}$ $(m=1,\ldots,N)$ are given by
\begin{equation}\label{eq:dm}
d_{m}^{\dagger}=\sum_{l=1}^{N}\left(c_{l}^{\dagger}V_{lm}+c_{l}U_{lm}\right),
\end{equation}
and the BdG Hamiltonian can be written in terms of quasihole (and/or quasiparticle) operators as follows:
\begin{equation}\label{eq:BdGd}
\begin{split}
\mathcal{H}_{\text{BdG}}&=-\sum_{m=1}^{N}\varepsilon_{m}d_m^{\dagger}d_m + \frac{1}{2}\sum_{m=1}^{N}\varepsilon_{m} + \frac{1}{2}\text{tr}(t).
\end{split}
\end{equation}

The ground state of the BdG Hamiltonian, $|\Psi_{0}\rangle$, can be represented as paired fermions and has the form of
\begin{equation}\label{eq:Psi0}
|\Psi_{0}\rangle\equiv\exp\left(\sum_{kl}g_{kl}c_{k}^{\dagger}c_{l}^{\dagger}\right)|0\rangle_{c},
\end{equation}
where $g_{kl}=-g_{lk}$ is the pairing function and $|0\rangle_{c}$ is the vacuum state of fermions, i.e., $c_{l}|0\rangle_{c}=0$ for $l=1,\ldots,N$. On the other hand, $|\Psi_{0}\rangle$ must be the vacuum of the Bogoliubov quasiparticles, namely $d_m^{\dagger}|\Psi_{0}\rangle=0$ for $m=1,\ldots,N$. Here we emphasize that $d_{m}^{\dagger}$ is a quasihole creation operator and thereby a quasiparticle annihilation operator in the quasihole representation. It is easy to verify that the pairing function should be
\begin{equation}
g_{kl}=\frac{1}{2}(VU^{-1})_{kl}
\end{equation}
so that the BCS state in Eq.~\eqref{eq:Psi0} is annihilated by all $d_{m}^{\dagger}$'s. However, it is noted that Eq.~\eqref{eq:Psi0} is valid for a system with even fermion parity only, while for a system with odd fermion parity, unpaired fermions have to be involved. Therefore it is more convenient to construct the ground state $|\Psi_{0}\rangle$ by filling all the quasihole states,
\begin{equation}\label{eq:Psi0d}
|\Psi_0\rangle = \prod_{m=1}^{N}d_{m}^{\dagger}|0\rangle_{d},
\end{equation}
where $|0\rangle_{d}$ is the vacuum of Bogoliubov quasiholes, i.e., $d_{m}|0\rangle_{d}=0$ for $m=1,\ldots,N$.
Notice that all the states in the Fock space of quasiparticles (and/or quasiholes) will be annihilated by the fully filling operator $\prod_{m=1}^{N}d_{m}^{\dagger}$ except the vacuum of quasiholes $|0\rangle_{d}$, because $\left(d_{m}^{\dagger}\right)^2=0$.
Thus the initial state $|0\rangle_{d}$ can be replaced by a simple direct-product state $|\cdots\rangle_{c}$ (in the basis of original fermions) as long as $|\cdots\rangle_{c}$ has the same fermion parity as $|0\rangle_{d}$. Otherwise, we have $\prod_{m=1}^{N}d_{m}^{\dagger}|\cdots\rangle_{c}=0$. In particular, we have
\begin{equation}\label{eq:Psi0c}
|\Psi_0\rangle = \prod_{m=1}^{N}d_{m}^{\dagger}|0\rangle_{c}
\end{equation}
in the presence of even fermion number parity.

A Gutzwiller projected paired state of fermions is obtained by removing all the components consisting of empty or multioccupied sites and can be expressed as
\begin{equation}\label{eq:Psi}
|\Psi\rangle = P_{G}|\Psi_{0}\rangle,
\end{equation}
where $P_{G}$ is the Gutzwiller projector which imposes the single-occupancy condition. Such a Gutzwiller projected state is widely used as a trial wave function for quantum spin systems. In the next section, we will demonstrate that both unprojected and projected paired states of fermions can be expressed as tensor-network states by using the MPO-MPS construction.

\section{Tensor-network representation of paired states of fermions: MPO-MPS method}\label{sec:mpo-mps}

A key observation was made in Ref.~[\onlinecite{wu2019}] that a single-particle creation/annihilation operator can be rewritten as an MPO. This can be naturally generalized to the Bogoliubov quasiparticle/quasihole creation/annihilation operators. To be explicit, the quasihole creation operator $d_{m}^{\dagger}$ can be written as an MPO with bond dimension $D=2$ as follows:
\begin{equation}
d_{m}^{\dagger} =
\left(\begin{array}{cc}
0 & 1
\end{array}\right)\left[\prod_{l=1}^{N}
\left(\begin{array}{cc}
1 & 0\\\
V_{lm}c_{l}^\dagger+U_{lm}c_{l} & 1
\end{array}\right)\right]
\left(\begin{array}{cc}
1 \\
0
\end{array}\right).\label{eq:MPO_dm}
\end{equation}
Notice that dummy column and row are employed in Eq.~\eqref{eq:MPO_dm} to ensure the open boundary condition of MPO.

\begin{figure}[tbp]
	\includegraphics[width=8.4cm]{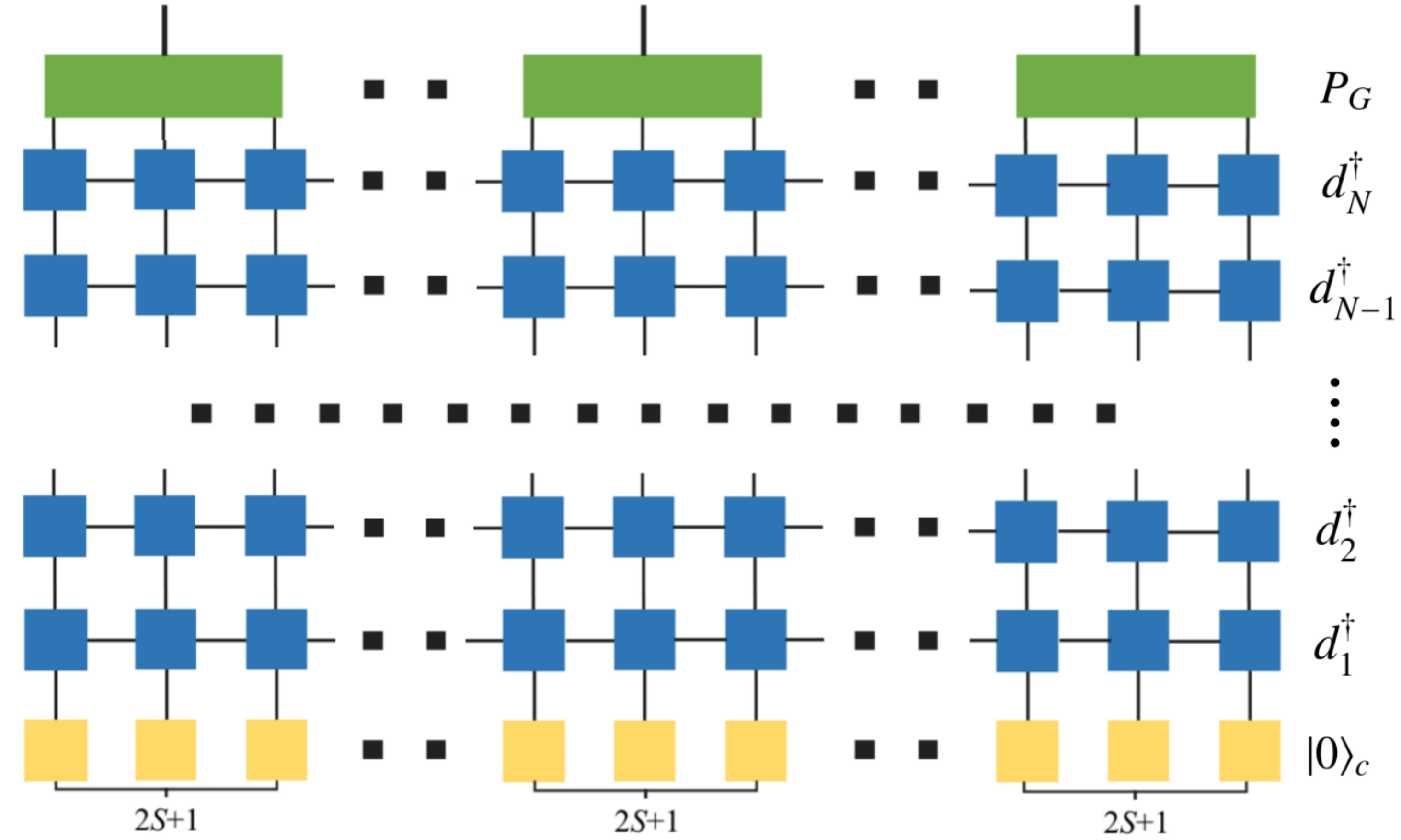}
	\caption{Schematics of the MPO-MPS method based on Eqs.~\eqref{eq:Psi0c} and \eqref{eq:Psi}. Here $2S+1$ species of fermions are introduced to represent a quantum spin-$S$. }\label{fig:algorithm}
\end{figure}

Based on the MPO expression in Eq.~\eqref{eq:MPO_dm}, the unprojected state $|\Psi_{0}\rangle$ given in Eq.~\eqref{eq:Psi0} and Eq.~\eqref{eq:Psi0c} and the Gutzwiller projected state $|\Psi\rangle$ given in Eq.~\eqref{eq:Psi} can be converted into an MPS form. The procedure of the MPO-MPS method is illustrated in Fig.~\ref{fig:algorithm}, and is made of two or three steps as follows:

(1) Initialize the vacuum state $|0\rangle_{c}$ as an MPS with bond dimension $D=1$.

(2) Act the $N$ MPOs in accordance with the Bogoliubov quasiholes $d^{\dagger}_m$'s iteratively onto the vacuum MPS of $|0\rangle_{c}$. Each action generates one new MPS, and such a new MPS should be compressed with the help of the so-called mixed canonical form of MPS~\cite{schollwock11}, by using singular value decomposition (SVD) at every intermediate step. Otherwise, the bond dimension of the obtained MPS will grow exponentially with the number of Bogoliubov quasiholes, i.e., $D=2^{N}$. So far we have obtained an MPS for the unprojected paired state of fermions, $|\Psi_{0}\rangle$.

(3) To obtain the projected state $|\Psi\rangle$, we apply the Gutzwiller projector $P_G$ to the MPS $|\Psi_{0}\rangle$ obtained in step (2).
Below we shall demonstrate how the Gutzwiller projection is implemented and show that $|\Psi\rangle$ is an MPS, too.

{\em  Implementation of Gutzwiller projection. ---} For a spin-$S$ system with $L$ lattice sites, it is convenient to use the interleaved site index, $l=1,\ldots,(2S+1)L$, instead of the original lattice site index $j=1,\ldots,L$ and spin/flavor index $\alpha=-S,\ldots,S$, such that the Gutzwiller projection is implemented on $(2S+1)$ neighboring interleaved sites. Note that the interleaved site index is related to lattice site and spin/flavor indices as $l=(j,\alpha)$.

To illustrate the Gutzwiller projection, we follow Refs.~[\onlinecite{liu2010,liu2010_2,liu2012}] to introduce $2S+1$ species of fermionic parton operator (Abrikosov fermion) $c_{j\alpha}^{\dagger}$ and write the three components of spin, $S_j^{a}$ ($a=x, y, z$), in terms of these partons as follows:
\begin{equation}\label{eq:spin-S}
S_{j}^{a} = \sum_{\alpha\beta}c_{j\alpha}^{\dagger}I^{a}_{\alpha\beta}c_{j\beta}
\end{equation}
with the single-occupancy constraint imposed at each lattice site $j$,
\begin{equation}\label{eq:constraint}
\sum_{\alpha}c_{j\alpha}^{\dagger}c_{j\alpha}=1,
\end{equation}
where $\alpha$ and $\beta$ are spin/flavor indices and $I^{a}$ is the spin-$S$ matrix representation of spin operator $S^{a}$.

At the end of step (2), one has obtained an MPS for the unprojected paired state of fermions as follows:
\begin{equation}\label{eq:MPSPsi0}
|\Psi_{0}\rangle=\sum_{\bm{s}}\mathrm{sgn}(\bm{s})A^{s_{1}}[1]A^{s_{2}}[2]\cdots{}A^{s_{N-1}}[N-1]A^{s_{N}}[N]|\bm{s}\rangle,
\end{equation}
where $N=(2S+1)L$ is the number of interleaved lattice sites, $A^{s_{l}}[l]$ is the matrix associated with fermion occupation number $s_{l}$ at interleaved site $l$, and $\bm{s}=\otimes_{l=1}^{N}s_{l}$ characterizes the basis of the fermion Fock space together with the fermion sign function $\mathrm{sgn}(\bm{s})=\pm 1$. We found that the Gutzwiller projected state can be written as an MPS as well,
\begin{equation}\label{eq:MPSPsi}
|\Psi\rangle=\sum_{\bm{\tau}}\text{sgn}(\bm{\tau})B^{\tau_{1}}[1]B^{\tau_{2}}[2]\cdots{}B^{\tau_{L-1}}[L-1]B^{\tau_{L}}[L]|\bm{\tau}\rangle,
\end{equation}
where $\bm{\tau}=\otimes_{j=1}^{L}\tau_{j}$ denotes the basis of the spin Hilbert space, and the associated matrix $B^{\tau_{j}}$ is given by
\begin{equation}
B^{\tau_{j}} =\left\lbrace\begin{array}{ll}
\prod_{\alpha}A^{s_{l}}[l]|_{l=(j,\alpha)}, & \text{if } \sum_{\alpha}s_{l=(j,\alpha)}=1,\\
0, & \text{otherwise}.
\end{array}
\right.
\end{equation}
Here the interleaved site index $l=(j,\alpha)=j(2S+1)-S+\alpha$, the spin index $\alpha$ runs from $-S$ to $S$, so that $l$ runs from $1$ to $N=(2S+1)L$. Note that $\text{sgn}(\bm{\tau})$ in Eq.~\eqref{eq:MPSPsi} is well defined and is given by $\text{sgn}(\bm{\tau})=\text{sgn}(\bm{s})$ as long as the single-occupancy condition is satisfied. In practice, when one calculates the expectation value for an operator $\hat{O}$, the fermion sign $\text{sgn}(\bm{\tau})$ and/or $\text{sgn}(\bm{s})$ can be absorbed in the operator $\hat{O}$, then the MPS itself will be implemented as a bosonic MPS.

{\em  Maximally localized Wannier orbitals. ---} As mentioned, one has to compress the matrices $A^{s_{l}}[l]$ in Eq.~\eqref{eq:MPSPsi0} after the acting of each MPO $d^{\dagger}_{m}$ on the MPS, which is done by utilizing SVD. The truncation error is unavoidable during the process of compression of matrices.

To reduce the truncation error of MPS, we would like to follow Ref.~[\onlinecite{wu2019}] to exploit the idea of ``maximally localized Wannier orbitals"~\cite{mlwo1,mlwo2,mlwo3,mlwo4,mlwo5}. These single-particle orbitals have minimum spatial overlap with each other, which allows us to reduce the entanglement entropy when each MPO associated with one of these single-particle orbitals is applied. These maximally localized Wannier orbitals are determined as follows: First, we define position operators for particles and holes in interleaved lattice sites,
\begin{equation}
  \hat{X} = \sum_{l}lc_{l}^{\dagger}c_{l}, \qquad
  \hat{X}' = \sum_{l}lc_{l}c_{l}^{\dagger}.
\end{equation}
For the Bogoliubov quasiparticles/quasiholes, the position operators are projected into a matrix $\tilde{X}$ with matrix element
\begin{equation}\label{eq:Xmn}
\tilde{X}_{mn}={}_{c}\langle{}0|d_m\hat{X}d_{n}^{\dagger}|0\rangle_{c} + {}_{c}\langle{}1|d_m\hat{X}'d_{n}^{\dagger}|1\rangle_{c},
\end{equation}
where $|1\rangle_{c}$ is the fully occupied state of fermions, i.e., $c^\dag_l|1\rangle_{c}=0$ for $l=1,\ldots,N$.
Note that the particle-hole symmetric form in Eq.~\eqref{eq:Xmn} provides an unambiguous definition even in the limit of vanishing pairing strength. Second, the matrix $\tilde{X}$ can be diagonalized by an $SU(N)$ matrix $W$, i.e., $W^{\dagger}\tilde{X}W=\mathrm{diag}\{x_{1},\cdots,x_{N}\}$, where the eigenvalues $x_{n}$ are sorted as $x_{1}<\cdots<x_{N}$. Thus we find out a set of single-particle operators $\{f^{\dagger}_{n}\}$ associated with the eigenvalues $\{x_{n}\}$ as follows:
\begin{equation}
f_{n}^{\dagger} = \sum_{m}d_{m}^{\dagger}W_{mn} = \sum_{l}\left[c_{l}^{\dagger}(VW)_{ln} + c_{l}(UW)_{ln}\right].
\end{equation}
Since $W$ is an $SU(N)$ matrix, we have
\begin{equation*}
\prod_{n=1}^{N}f_{n}^{\dagger}=\prod_{m=1}^{N}d_{m}^{\dagger}.
\end{equation*}
Therefore the paired state $|\Psi_{0}\rangle$ given in Eq.~\eqref{eq:Psi0c} can be rewritten in terms of the single-particle operators $\{f_{n}^{\dagger}\}$,
\begin{equation}\label{eq:Psi0c_f}
|\Psi_0\rangle = \prod_{n=1}^{N}f_{n}^{\dagger}|0\rangle_{c}.
\end{equation}

\begin{figure*}[tbp]
	\includegraphics[width=17cm]{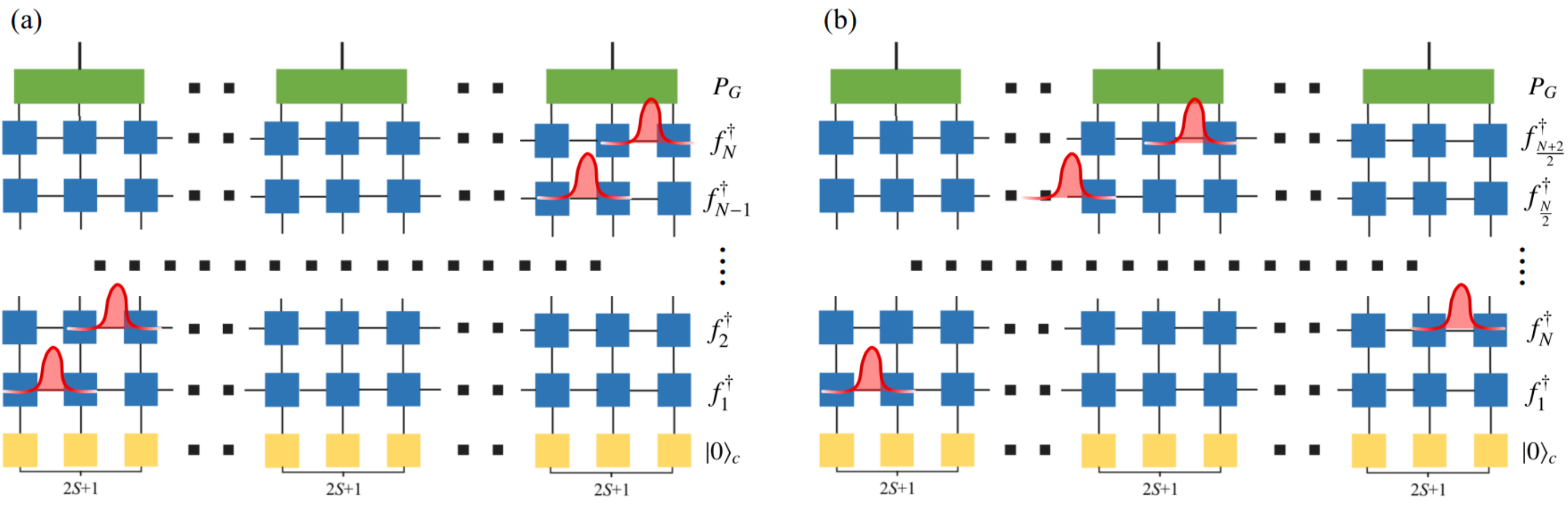}
	\caption{{\color{black}Schematics of the MPO-MPS method based on Eqs.~\eqref{eq:Psi0c_f} and \eqref{eq:Psi} with the (a) left-to-right scheme and{black} left-meet-right scheme. The red wave packet denotes the leading component of a maximally localized Wannier orbital $f^\dagger_l$.
	The $2S+1$ species of fermions are introduced to represent a quantum spin-$S$.} }\label{fig:MPOorder}
\end{figure*}

When acting the MPO
\begin{equation}
f_{n}^{\dagger} =
\left(\begin{array}{cc}
0 & 1
\end{array}\right)\left[\prod_{l=1}^{N}
\left(\begin{array}{cc}
1 & 0\\\
c_{l}^{\dagger}(VW)_{ln} + c_{l}(UW)_{ln} & 1
\end{array}\right)\right]
\left(\begin{array}{cc}
1 \\
0
\end{array}\right)\label{eq:MPO_fn}
\end{equation}
on the MPS, the matrices $A^{s_{l}}[l]$ will change considerably only when the interleaved site $l$ is near the position $x_{n}$, because the single-particle wave functions associated with $f_{n}^{\dagger}$ are maximally localized thereby separated from one another. On the other hand, different orderings of $f^{\dagger}_{n}$ only differ by a global factor $\pm 1$ in $|\Psi_0\rangle$. So that one can act $f^{\dagger}_{n}$ in $|\Psi_0\rangle$ by the ordering of {\color{black} either ``left-to-right'' [see Fig.~\ref{fig:MPOorder}(a)] or ``left-meet-right" which starts from the left or right edge and gradually moves toward the center [see Fig.~\ref{fig:MPOorder}(b)]}. This procedure drastically minimizes the truncation error.

{\color{black} It is worth mentioning that the idea of minimizing entanglement and truncation errors by optimizing single-particle orbitals and choosing proper ordering of the action of MPOs have also been used in other contexts of tensor networks~\cite{legeza03,legeza03_2,murg10,murg12bethe,krumnow16,pastori19}}.

{\em  Another MPO-MPS representation of paired states. ---} As mentioned in Ref.~[\onlinecite{wu2019}], Eq.~\eqref{eq:Psi0} has another MPO-MPS representation which can be achieved by rewriting Eq.~\eqref{eq:Psi0} as
\begin{equation}\label{eq:Psi0Wk}
|\Psi_{0}\rangle=\prod_{kl}(1+g_{kl}c_{k}^\dagger{}c_{l}^\dagger)|0\rangle_{c}=\prod_{k}\hat{W}_{k}|0\rangle_{c},
\end{equation}
where
\begin{equation}\label{eq:MPO_Wk}
\begin{split}
\hat{W}_{k} =
\left(\begin{array}{cc}
1 & 0
\end{array}\right)
&\left[\prod_{l=1}^{k-1}
\left(\begin{array}{cc}
1 & g_{kl}c^{\dagger}_l\\\
0 & 1
\end{array}\right)\right]
\left(\begin{array}{cc}
1 & c^{\dagger}_k\\\
-c^{\dagger}_k & 1
\end{array}\right) \\&
\times\left[\prod_{l=k+1}^{N}
\left(\begin{array}{cc}
1 & 0\\\
g_{kl}c^{\dagger}_l & 1
\end{array}\right)\right]
\left(\begin{array}{cc}
1 \\
0
\end{array}\right)
\end{split}
\end{equation}
is also an MPO with bond dimension $D=2$. However, the formalism based on Eqs.~\eqref{eq:Psi0Wk} and \eqref{eq:MPO_Wk} is hard to be improved with the help of maximally localized Wannier orbitals.
Moreover, the numerical calculation will become unstable when the matrix $U$ thereby the pairing function $g_{kl}$ become singular.
{\color{black} Later we will show that the MPO-MPS method based on Eq.~\eqref{eq:Psi0Wk} fails
when the pairing amplitude is very small, but the one based on Eq.~\eqref{eq:Psi0c} still works well.}
We shall examine and compare these different MPO-MPS representations [in accordance with Eqs.~\eqref{eq:Psi0c}, \eqref{eq:Psi0c_f}, and \eqref{eq:Psi0Wk} respectively] in the next section.

{\em  ``Bosonization" of fermionic states. ---}
Since the fermionic sign brings extra complications in the MPO-MPS formalism, we would like to bosonize the fermionic tensor networks (MPOs and MPSs) with the help of Jordan-Wigner transformation,
\begin{equation}\label{eq:JW}
c^{\dagger}_{l} = \left[\prod_{k=1}^{l-1}\sigma_{k}^{z}\right]\sigma^{+}_{l}, \quad c_{l} = \left[\prod_{k=1}^{l-1}\sigma_{k}^{z}\right]\sigma^{-}_{l}.
\end{equation}
Thus the quasihole operator $d^{\dagger}_{m}$ defined in Eq.~\eqref{eq:dm} and the corresponding MPO form given in Eq.~\eqref{eq:MPO_dm} be rewritten in terms of pseudospin-1/2 as follows:
\begin{equation}
d_{m}^{\dagger} =
\left(\begin{array}{cc}
0 & 1
\end{array}\right)\left[\prod_{l=1}^{N}
\left(\begin{array}{cc}
1 & 0\\\
V_{lm}\sigma_{l}^{+}+U_{lm}\sigma^{-}_{l} & \sigma^{z}_{l}
\end{array}\right)\right]
\left(\begin{array}{cc}
1 \\
0
\end{array}\right).\label{eq:MPO_dm2}
\end{equation}
Note that this pseudospin MPO expression can be used for the maximally localized Wannier orbitals in Eq.~\eqref{eq:MPO_fn} as well.

\section{Transverse field XY model: a benchmark}\label{sec:TFXY}

In this section, we shall study the 1D transverse field XY (TFXY) model, which is a spin-1/2 model defined by the following Hamiltonian:
\begin{equation}\label{eq:xy}
H_{XY }= \sum_{j=1}^{L} \left( J_{x}\sigma^{x}_{j}\sigma^{x}_{j+1}+ J_{y}\sigma^{y}_{j}\sigma^{y}_{j+1}+h_{z}\sigma^{z}_{j} \right),
\end{equation}
where the periodic boundary condition is imposed by $\sigma^{x,y,z}_{L+1}=\sigma^{x,y,z}_{1}$. This model is exactly solvable and will be a good benchmark of our MPO-MPS method. We will compare various MPO-MPS expressions in accordance with Eqs.~\eqref{eq:Psi0c}, \eqref{eq:Psi0c_f}, and \eqref{eq:Psi0Wk}.

The 1D TFXY model can be fermionized by the inverse Jordan-Wigner transformation. The resulting spinless fermion model reads
\begin{equation}\label{eq:xy_fermion}
\begin{split}
H_{XY}=\sum_{j=1}^{L-1}&\left(J_{-}c_{j}c_{j+1}-J_{+}c_{j}^{\dagger}c_{j+1}+\mathrm{h.c}\right) \\+\sum_{j=1}^{L}&h_{z}(2c_{j}^{\dagger}c_{j}-1) + h_{{L,1}},
\end{split}
\end{equation}
where $J_{\pm}=J_{x}\pm{}J_{y}$, and the boundary term reads $$h_{{L,1}}=\left(J_{+}c_{L}^{\dagger}c_{1}-J_{-}c_{L}c_{1}+ \mathrm{h.c}\right)e^{i\pi{}\hat{N}}.$$ Here $\hat{N}=\sum_{j=1}^{L}c_{j}^{\dagger}c_{j}$ is the total fermion number.
Since the total fermion parity $e^{i\pi{}\hat{N}}=\pm{}1$ is a good quantum number, the fermionic Hamiltonian will become quadratic when the eigenvalue of $e^{i\pi{}\hat{N}}$ is fixed. Thus, we are able to obtain the {\em  exact} ground-state energy $\varepsilon_{XY}$ and all the eigenstates of $H_{XY}$ by the Bogoliubov transformation.

Now let us examine how efficient the ground state $|\Psi_{XY}\rangle$ can be computed by the MPO-MPS method, which is an unprojected paired state of fermions and will be computed by the maximally localized Wannier orbitals $\{f^{\dagger}_l\}$ [by using Eq.~\eqref{eq:Psi0c_f}], original Bogoliubov quasiholes $\{d^{\dagger}_m\}$ [by using Eq.~\eqref{eq:Psi0c}], and the pairing function $\{W_{k}\}$ [by using Eq.~\eqref{eq:Psi0Wk}], respectively. To see the precision of these MPO-MPS methods, we define the energy deviation per site,
\begin{equation}\label{eq:dExy}
\delta{}\varepsilon_{XY} = \frac{1}{L}\left(\langle\Psi_{XY}|{H}_{XY}|\Psi_{XY}\rangle - \varepsilon_{XY}\right),
\end{equation}
where $\varepsilon_{XY}$ is the exact ground-state energy obtained by the Bogoliubov transformation and $|\Psi_{XY}\rangle$ is computed by MPO-MPS methods. In order to monitor the precision after each MPO is applied and the truncation is done, we divide the spin chain into two parts (denoted by $A$ and $B$) from the middle and calculate the von Neumann entanglement entropy of the reduced density matrix for $A$,
\begin{equation}\label{eq:Sc}
S_{c} = -\text{tr}\left(\rho_{A}\log \rho_{A}\right),
\end{equation}
where $\rho_{A}=\text{Tr}_{B}\rho=\text{Tr}_{B}|\Psi\rangle\langle\Psi|$, and $|\Psi\rangle$ is measured after every MPO is applied.

\begin{figure*}[tbp]
	\includegraphics[width=14.8cm]{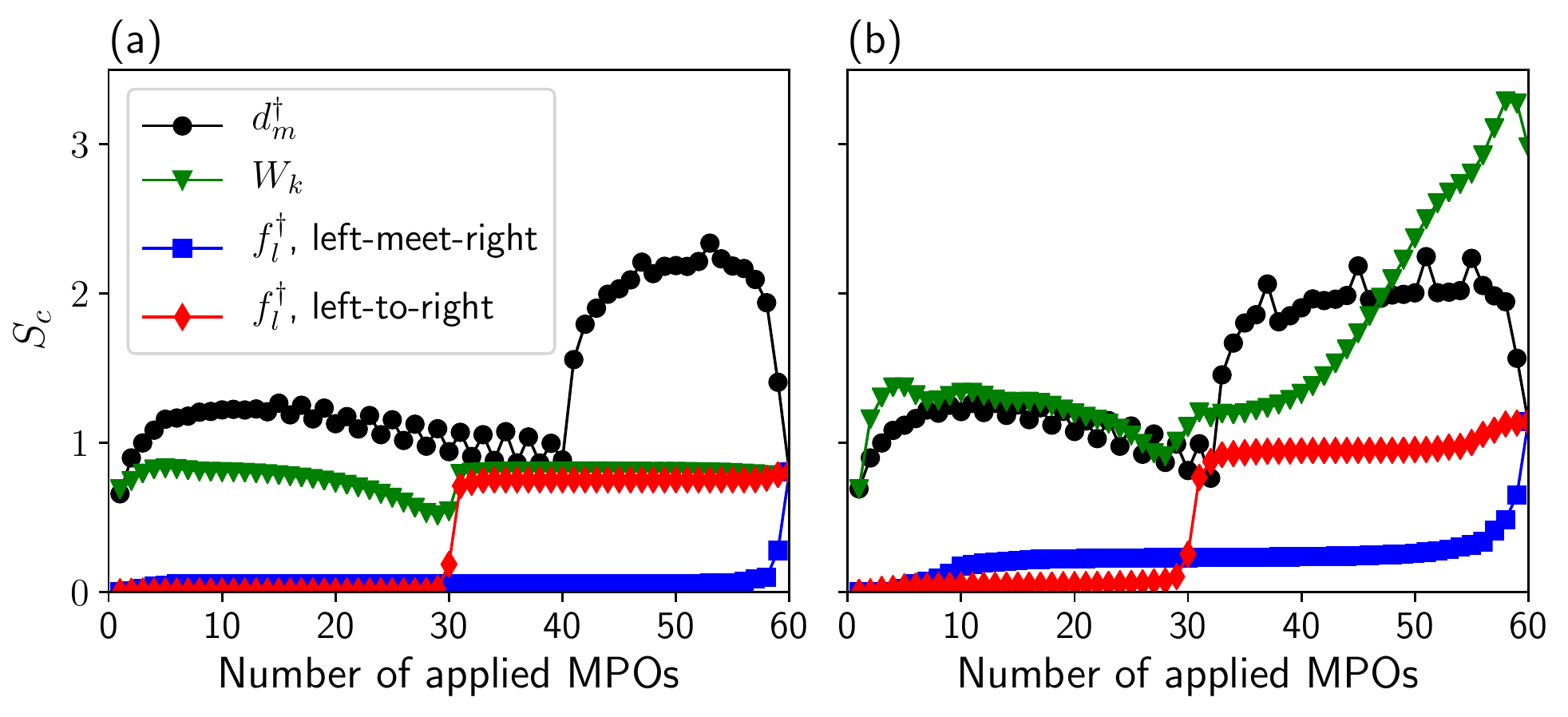}
	\caption{The entanglement entropy $S_{c}$ [defined in Eq.~\eqref{eq:Sc}] versus the number of applied MPOs, where $d^{\dagger}_m$ denotes the original Bogoliubov quasiparticles, $f^{\dagger}_l$ the maximally localized Wannier orbitals, and $W_{k}$ is defined in Eq.~\eqref{eq:MPO_Wk}. We set $L=60$, $D=200$, $J_x=1$, $h_z=1$, and (a) $J_y=0.5$ and{black} $J_y=0.8$.}\label{fig:wannier}
\end{figure*}

The deviation of the ground-state energy $\delta{}\varepsilon_{XY}$ is given in Table~\ref{tab:XY}, and the entanglement entropy $S_{c}$ versus the number of applied MPOs is plotted in Fig.~\ref{fig:wannier}. It turns out that the $f_{l}^{\dagger}$-MPO-MPS method using maximally localized Wannier orbitals and the ``left-meet-right'' scheme gives rise to very accurate ground states. Namely, $\delta{}\varepsilon_{XY}$ is always less than $10^{-10}$ for all the model parameters chosen, i.e., $J_{y}/J_{x}$ and $h_{z}/J_{x}$, where we fix $J_{x}=1$. It can be seen from Fig.~\ref{fig:wannier} that $S_{c}$ increases very slowly and keeps a small value until the last few MPOs are applied. So that the truncation error will keep a small value when the MPSs are compressed after each MPO is applied.

As the comparison to the $f_{l}^{\dagger}$-MPO-MPS method, the $d_{m}^{\dagger}$-MPO-MPS method using original Bogoliubov quasiholes is also investigated. For this case, the ordering for the action of $d_{m}^{\dagger}$-MPOs is from low to high in their corresponding single-particle energies. It gives rise to rather reasonable results with $\delta{}\varepsilon_{XY}\sim 10^{-3}$-$10^{-8}$, although the precision is much poorer than the $f_{l}^{\dagger}$-MPO method (see Table~\ref{tab:XY}). Finally, the $W_{k}$-MPO-MPS method (also with a ``left-to-right'' scheme) is found to be unstable and fail when the pairing function $g_{kl}$ becomes singular, e.g., at $(J_{y}/J_{x},h_{z}/J_{x})=(0.8,0.5)$ and $(J_{y}/J_{x},h_{z}/J_{x})=(0.8,1.0)$ (see Table~\ref{tab:XY}).

\begin{table}[tbp]
\renewcommand\arraystretch{1.5}
\setlength\tabcolsep{0.4cm}
\begin{tabular}{c|ccc}
	\hline
	\hline
	\diagbox{$(J_y,h_{z})$}{MPO} & $f^{\dagger}_l$ &  $d^{\dagger}_m$  &  $W_k$\\
	\hline
	(0.5, 0.5) & $<10^{-12}$ & $<10^{-3}$ & $<10^{-6}$ \\
	(0.8, 0.5) & $<10^{-10}$ & $<10^{-3}$ & 0.58\\
	(0.5, 1.0) & $<10^{-14}$ & $<10^{-5}$ & $<10^{-10}$ \\
	(0.8, 1.0) & $<10^{-10}$ & $<10^{-4}$ & 0.22\\
	(0.5, 1.5) & $<10^{-14}$ & $<10^{-8}$ & $<10^{-13}$ \\
	(0.8, 1.5) & $<10^{-11}$ & $<10^{-7}$ & $<10^{-4}$\\
	\hline
\end{tabular}
\caption{The energy deviation $\delta{}\varepsilon_{XY}$ defined in Eq.~\eqref{eq:dExy} and evaluated by three MPO-MPS methods. Here $f^{\dagger}_l$ denotes the method using maximally localized Wannier orbitals and the ``left-meet-right'' scheme; $d^{\dagger}_m$ denotes the method using original Bogoliubov quasiparticles (ordered with respect to single-particle energies); and $W_k$ means the MPO-MPS method using the pairing function $g_{kl}$ and defined in Eqs.~\eqref{eq:Psi0Wk} and \eqref{eq:MPO_Wk} (with the ``left-to-right'' scheme). We choose lattice size $L=60$, bond dimension $D=200$, and set $J_{x}=1$ in the Hamiltonian $H_{XY}$.}\label{tab:XY}
\end{table}

\section{$SO(3)$-symmetric $S=1$ spin chains}\label{sec:bbq}

In this section, we shall study $SO(3)$ rotationally invariant spin-1 chains~\cite{bbq1,bbq2,bbq3,bbq1d0,bbq1d1,bbq1d2,bbq1d3,bbq1d4,bbq1d5} by the $f_{l}^{\dagger}$-MPO-MPS method, say, using maximally localized Wannier orbitals and the ``left-meet-right'' scheme. The Hamiltonian for these $S=1$ spin chains is of the following bilinear-biquadratic (BBQ) form:
\begin{equation}\label{eq:BBQ}
H_{\scriptsize{\text{BBQ}}}=\sum_{j=1}^{L}\left[J\bm{S}_{j}\cdot\bm{S}_{j+1}+K\left(\bm{S}_{j}\cdot\bm{S}_{j+1}\right)^{2}\right],
\end{equation}
where $\bm{S}_{j}$ is the $S=1$ spin operator at the $j$th site and $L$ is the length of the spin chain. The periodic boundary condition is imposed by $\bm{S}_{L+1}=\bm{S}_{1}$.
Note that the Hamiltonian in Eq.~\eqref{eq:BBQ} can be reparameterized by setting $J=\cos\theta$ and $K=\sin\theta$ apart from an unimportant factor, such that $$H_{\scriptsize{\text{BBQ}}}=\sum\left[\cos\theta{}\bm{S}_{j}\cdot\bm{S}_{j+1}+\sin\theta{}\left(\bm{S}_{j}\cdot\bm{S}_{j+1}\right)^{2}\right].$$
The phase diagram of the $SO(3)$-symmetric spin-1 chain is well studied:
(1) for $\pi/2<\theta<5\pi/4$, the system is in a gapless ferromagnetic phase;
(2) for $\pi/4\leq\theta\leq\pi/2$, it is a critical phase~\cite{bbq1d2,bbq1d4,bbq3}, which includes an exactly solvable Uimin-Lai-Sutherland (ULS) point~\cite{bbq1d0,SU3ULSL,SU3ULSS} at $\theta=\pi/4$ ($J=K>0$);
(3) for $-\pi/4<\theta<\pi/4$, it is the gapped Haldane phase containing the Affleck-Kennedy-Lieb-Tasaki (AKLT) point~\cite{aklt} at $\theta=\tan^{-1}(1/3)$;
(4) for  $-3\pi/4<\theta<-\pi/4$, it is a dimerized phase;
and (5) there exists an exactly solvable Takhtajan-Babujian (TB) critical point~\cite{TB0,TB1} separating the Haldane and the dimerized phase at $\theta=-\pi/4$ ($J=-K>0$).

It was revealed in Ref.~[\onlinecite{liu2012}] that the Gutzwiller projected wave functions of paired fermions are very good trial wave functions for the ground states of the antiferromagnetic BBQ model in the regime $K <J\,\,(-3\pi/4 < \theta < \pi/4)$, where variational energies and static spin correlation functions were calculated by using the VMC method. In this section, we will demonstrate that such trial wave functions can be efficiently converted into MPSs by the MPO-MPS method.
{\color{black} To check how good the MPS approximation is, we benchmark the truncation error, the ground-state energy, and the entanglement entropy.} Moreover, we find that the dynamic spin correlation (spin spectral function) can be easily evaluated, since the MPO-MPS method can be naturally cooperated with the Chebyshev kernel polynomial method to compute spectral functions~\cite{chebyshevmps}.

\subsection{Fermionic theory, trial wave function, and spin spectral function}
To formulate the fermionic theory and derive the trial wave function for a spin $S=1$ system, we follow Ref.~[\onlinecite{liu2010}] to introduce three species of fermions: $c_{1}$, $c_{0}$, $c_{-1}$. Then the spin operators can be represented in terms of these fermions as in Eq.~\eqref{eq:spin-S}. To see the $SO(3)$ spin rotational symmetry, it is more convenient to use the Cartesian basis: $c_x=i (c_1-c_{-1})/\sqrt{2}$, $c_y=(c_1+c_{-1})/\sqrt{2}$, and $c_z=-ic_0$, and define two $SO(3)$ invariant bond operators $\hat{\chi}_{ij}$ and $\hat{\Delta}_{ij}$ as follows:
\begin{equation}
\hat{\chi}_{ij}=\sum_{\alpha=x,y,z}c^{\dagger}_{i\alpha}c_{j\alpha},\quad\hat{\Delta}_{ij}=-\sum_{\alpha=x,y,z}c_{i\alpha}c_{j\alpha}.
\end{equation}
Thus, the $SO(3)$ symmetric model given in Eq.~\eqref{eq:BBQ} can be rewritten in terms of $\hat{\chi}_{ij}$ and $\hat{\Delta}_{ij}$,
\begin{equation}\label{eq:HBBQ2}
H_{\text{BBQ}}=-\sum_{j=1}^{L}\left[J{}\hat{\chi}_{j,j+1}^{\dagger}\hat{\chi}_{j,j+1}+(J-K)\hat{\Delta}_{j,j+1}^{\dagger}\hat{\Delta}_{j,j+1}\right].
\end{equation}
It is expected that paired states of fermions with $\langle\hat{\Delta}_{j,j+1}\rangle\neq 0$ will be energetically favored when $J>K$ due to the last term in Eq.~\eqref{eq:HBBQ2}.

{\em  Trial wave function. ---} At the mean-field level, the $SO(3)$ symmetric Hamiltonian given in Eq.~\eqref{eq:HBBQ2} can be naturally decoupled to three copies of Kitaev's Majorana chains~\cite{kitaev1,kitaev2}:
\begin{equation}\label{eq:BBQmf2}
\begin{split}
&H_{\text{MF}}\equiv\sum_{\alpha=x,y,z}H_{\text{K}}^{(\alpha)},\\
&H_{\text{K}}^{(\alpha)}=\sum_{j=1}^{L}\left[-\chi{}c_{j\alpha}^{\dagger}c_{j+1\alpha} +\Delta{}c_{j\alpha}c_{j+1\alpha}+\mathrm{h.c.}\right]+ \lambda{}\sum_{j=1}^{L}c_{j\alpha}^\dagger{}c_{j\alpha}.
\end{split}
\end{equation}
Here $\chi$ and $\Delta$ are two mean-field order parameters, and $\lambda$ serves as the Lagrange multiplier to impose the particle number constraint given in Eq.~\eqref{eq:constraint} on average. For a given set of $\{\chi,\Delta,\lambda\}$, a mean-field ground state $|\Psi_{\text{MF}}(\chi,\Delta,\lambda)\rangle$ can be obtained. Thus the Gutzwiller projected wave function
\begin{equation}
|\Psi_{\text{BBQ}}(\chi,\Delta,\lambda)\rangle = P_{G}|\Psi_{\text{MF}}(\chi,\Delta,\lambda)\rangle
\end{equation}
can be treated as a trial wave function for the spin Hamiltonian $H_{\text{BBQ}}$ defined in Eq.~\eqref{eq:BBQ}, where $\{\chi,\Delta,\lambda\}$ is a set of variational parameters. Then the ground state can be obtained by minimizing the energy (per site),
\begin{equation}
E_g = \frac{1}{L}\frac{\langle\Psi_{\text{BBQ}}|H_{\text{BBQ}}|\Psi_{\text{BBQ}}\rangle}{\langle\Psi_{\text{BBQ}}|\Psi_{\text{BBQ}}\rangle}.
\end{equation}
It is worth mentioning that there are only two independent variational parameters $\Delta/\chi$ and $\lambda/\chi$ to determine the ground state. Here one subtlety is that for certain phases, e.g., Haldane phase, which is characterized as a $p$-wave weakly pairing state~\cite{liu2012}, it is desirable to use antiperiodic boundary conditions ($c_{L+1,\alpha} = -c_{1,\alpha}$) for the mean-field Hamiltonian in Eq.~\eqref{eq:BBQmf2} in order to obtain a nonvanishing Gutzwiller projected wave function.

With the help of the trial wave function $|\Psi_{\text{BBQ}}(\chi,\Delta,\lambda)\rangle$, the ground state and elementary excitations of the $SO(3)$ Hamiltonian $H_{\text{BBQ}}$ were studied by using VMC in Ref.~[\onlinecite{liu2012}] and Ref.~[\onlinecite{liu2014}], respectively. It was found that the Gutzwiller projected wave function is in surprisingly good agreement with the known result given by exact solution and/or DMRG when $J>K$. Below we will demonstrate that the MPO-MPS method provides an alternative and efficient way to perform calculations based on the Gutzwiller projected wave function. Furthermore, the spin spectral function can be computed by the combination of the MPO-MPS method and the Chebyshev kernel polynomial method.

{\em  Spin spectral function.---} The spin spectral function in $(\bm{q},\omega)$ space, which can be measured by the inelastic neutron scattering, is defined as
\begin{equation}\label{eq:Sqw}
S(\bm{q},\omega)=\sum_{\alpha=x,y,z}\langle{}0|S ^{\alpha}(\bm{q})\delta(\omega-\hat{H}-E_0)S^{\alpha}(-\bm{q})|0\rangle.
\end{equation}
Here $|0\rangle$ denotes the ground state of a Hamiltonian $\hat{H}$, and $E_0$ is the corresponding ground-state energy. Usually, for a given Gutzwiller projected wave function, such a spectral function is difficult to calculate by the VMC method although static spin correlation functions can be done. By contrast, there exist a slice of MPS-based accurate approaches to calculate spectral functions, such as correction-vector method~\cite{cv0,cv1,cv2,cv3}, time-dependent DMRG~\cite{tdmrg0,tdmrg1,tdmrg2,tdmrg3,tdmrg4}, and Chebyshev MPS~\cite{chebyshevmps,rechebyshevmps}. In this paper, we utilize the Chebyshev MPS method~\cite{chebyshevmps}, of which the framework is to expand the $\delta$ function in Eq.~\eqref{eq:Sqw} in terms of Chebyshev polynomials. The details of the Chebyshev MPS method can be found in Appendix~\ref{app:chbmps}.

\subsection{Numerical results and analyses}

Numerically, we will focus on four representative points in the phase diagram: (1) AKLT point at $\theta=\tan^{-1}(1/3)$ (or $K/J=1/3$), (2) TB point at $\theta=-\pi/4$ (or $K/J=-1$), (3) ULS point at $\theta=\pi/4$ (or $K/J=1$), and (4) Heisenberg point at $\theta=0$ (or $K=0,J=1$). Two of them, AKLT and Heisenberg points, are gapped, while the other two, TB and ULS points, are gapless. We shall study the trial wave function $|\Psi_{\text{BBQ}}(\chi,\Delta,\lambda)\rangle$ at these four points and use the parameters $\{\Delta/\chi, \, \lambda/\chi\}$ optimized by the VMC in Ref.~[\onlinecite{liu2012}]. Note that the definition of the mean-field order parameters $\chi$ and $\Delta$ are different from those defined in Ref.~[\onlinecite{liu2012}] by a factor of $J$ and $(J-K)$, respectively.

\begin{figure}[tpb]
	\includegraphics[width=8.6cm]{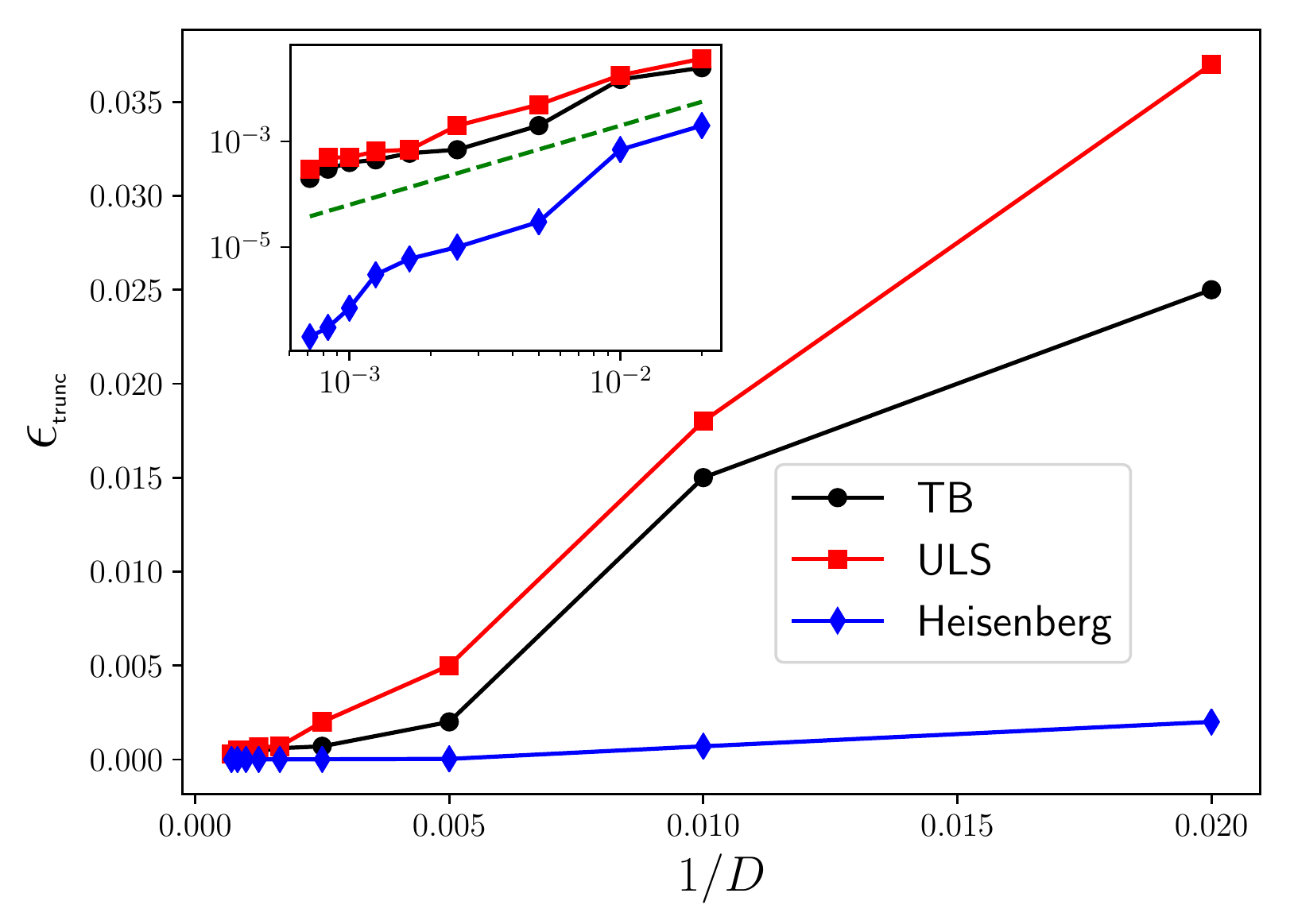}
	\caption{Truncation error $\epsilon_{\text{trunc}}$ versus the inverse bond dimension $1/D$. {\color{black} Inset: Log scale is used and the dashed line ($\epsilon_{\text{trunc}}\propto D^{-3/2}$) is a guide to the eye. The data are the same as those list in Table~\ref{tab:errL60}. } }\label{fig:truncerror}
\end{figure}

{\em  Truncation error. ---} To illustrate the precision of the MPO-MPS calculation, we introduce the truncation error $\epsilon_{\text{trunc}}$ of MPS that is truncated down to the leading $D$ singular values (more precisely, the upper bound of the truncation errors during the whole MPO-MPS process). Quantitatively, $\epsilon_{\text{trunc}}$ is defined as
\begin{equation}\label{eq:trunc_error}
\epsilon_{\mathrm{trunc}}=\sum_{j=1}^{N}\epsilon_{j}(D),
\end{equation}
where $\epsilon_{j}(D)$ is the sum of discarded squared singular values at the $j$th bond. The MPS truncation errors $\epsilon_{\mathrm{trunc}}$ for different bond dimension $D$ on an $L=60$ lattice are listed in Table~\ref{tab:errL60} and plotted in Fig.~\ref{fig:truncerror} for four points: AKLT, TB, ULS, and Heisenberg.

In general, we find that the MPO-MPS method works much more efficiently than the VMC for gapped states. For gapless states, it will achieve similar precision as the VMC for the same computing time. The reason is the following: The MPS truncation error $\epsilon_{\text{trunc}}$ decreases with the bond dimension $D$ nearly exponentially for gapped states, {\color{black} while it decreases in a power law for gapless states, $\epsilon_{\text{trunc}}\propto D^{-3/2}$.  The computing time for obtaining the MPS scales as $O(D^{3})$. So that the MPS truncation error $\epsilon_{\text{trunc}}$ is of order $O(t^{-1/2})$ for gapless states, where $t$ is the time consumption in the calculation. Meanwhile, the statistical error in the VMC is of $O(M^{-1/2})\propto O(t^{-1/2})$, where $M$ is the number of uncorrelated Monte Carlo measurements.}

\begin{table}[tpb]
	\renewcommand\arraystretch{1.5}
	\setlength\tabcolsep{0.3cm}
	\begin{tabular}{c|c|c|c|c}
		\hline
		\hline
		$D$ &  TB & ULS & Heisenberg & AKLT\\
		\hline
     10 &          &          &                      & $<10^{-15}$\\
		50 & $0.025$ & $0.037$ & $2\times{}10^{-3}$ & $<10^{-15}$\\
		100 & $0.015$ & $0.018$ & $7\times{}10^{-4}$ & $<10^{-15}$\\
		200 &  $8\times{}10^{-4}$ & $5\times{}10^{-3}$  & $3\times{}10^{-5}$ &  $<10^{-15}$\\
		400 &  $7\times{}10^{-4}$ & $2\times{}10^{-3}$  & $1\times{}10^{-5}$ &  $<10^{-15}$\\
		600 &   $6\times{}10^{-4}$ & $7\times{}10^{-4}$  & $6\times{}10^{-6}$ &  $<10^{-15}$\\
		800 &   $4.5\times{}10^{-4}$ & $6.5\times{}10^{-4}$  & $3\times{}10^{-6}$ &  $<10^{-15}$\\		
		1000 &  $4\times{}10^{-4}$ & $5\times{}10^{-4}$  & $7\times{}10^{-7}$ &  $<10^{-15}$\\
		1200 &  $3\times{}10^{-4}$ & $5\times{}10^{-4}$  & $3\times{}10^{-7}$ &  $<10^{-15}$\\
		1400 &  $2\times{}10^{-4}$ & $3\times{}10^{-4}$  & $2\times{}10^{-7}$ &  $<10^{-15}$\\
		\hline
	\end{tabular}
	\caption{Truncation errors $\epsilon_{\text{trunc}}(D)$ in the basis-optimized MPO-MPS process, which is defined in Eq.~\eqref{eq:trunc_error}. The lattice size is $L=60$.}\label{tab:errL60}
\end{table}

{\em  Ground-state energy and its variance. ---} The ground state energy and the energy variance have been computed by the basis-optimized MPO-MPS method and some results are list in Table~\ref{tab:EgBBQ}. Here the energy variance is defined as
\begin{equation}\label{eq:evar}
\varsigma\equiv\frac{1}{L}\sqrt{ \langle{}H_{\text{BBQ}}^{2}\rangle - \langle{}H_{\text{BBQ}}\rangle^{2}},
\end{equation}
which measures how the trial wave function $|\Psi_{\text{BBQ}}(\chi,\Delta,\lambda)\rangle$ deviates from an eigenstate of the Hamiltonian $H_{\text{BBQ}}$. Note that the energy variance $\varsigma$ defined in Eq.~\eqref{eq:evar} is hard to compute by VMC, which is different from the standard statistical deviation in VMC. We choose lattice size $L=60$ as in Table~\ref{tab:errL60} and Fig.~\ref{fig:truncerror} and use bond dimension $D=1400$ to obtain accurate values for gapless states (TB and ULS). However, smaller $D$ is sufficient to give the same precision for gapped states, namely, $D\ge 8$ for the AKLT point and $D\ge 100$ for the Heisenberg point.

\begin{table}[tpb]
\renewcommand\arraystretch{1.5}
\setlength\tabcolsep{0.1cm}
\begin{tabular}{c|ccccc}
	\hline
	\hline
	$K$ & $(\Delta/\chi,\lambda/\chi)$ & $E_g$ & $\varsigma$ & $E_g$(VMC) & $E^{*}_g$ \\
	\hline
	$\frac{1}{3}$\footnotemark[1] & $(1, 0)$ & $-\frac{2}{3}$ & $<10^{-15}$ & $-\frac{2}{3}$$\pm{}7\times{}10^{-15}$ & $-\frac{2}{3}$\\
	$-1$\footnotemark[2] & $(2.22, 2.0)$ & $-3.9909$ & $0.0389$ & $-3.9917\pm0.0012$ & $-4$ \\
	$1$\footnotemark[3] & $(0, 1)$ & $0.2996$ & $0.0150$ & $0.2997\pm0.0004$	
	 & $0.2971$ \\
	$0$\footnotemark[4] & $(0.98, 1.78)$ & $-1.4000$ & $0.0104$ & $-1.4001\pm{}0.0004$ & $-1.4015$ \\
	\hline
\end{tabular}

\footnotemark[1]{AKLT} 	
\footnotemark[2]{TB}
\footnotemark[3]{ULS}
\footnotemark[4]{Heisenberg}
\caption{Ground-state energy $E_{g}$ and the variance of energy $\varsigma$ calculated by the basis-optimized MPO-MPS method. The optimized mean-field parameters $(\Delta/\chi,\lambda/\chi)$ is given by the VMC in Ref.~[\onlinecite{liu2012}]. $E_g^{*}$ is the known result by exact solution (for AKLT~\cite{aklt}, TB~\cite{TB0,TB1} and ULS~\cite{bbq1d0,SU3ULSL,SU3ULSS}) or DMRG (for Heisenberg~\cite{S1dmrg}). Note that $E_g^{*}$ corresponds to per-site energies in the thermodynamic limit $L \rightarrow \infty$. We set $J=1$ and choose bond dimension $D=1400$ ($D=10$ for AKLT) and lattice size $L=60$. The ground state energy and its statistical deviation from the VMC calculation are also list for reference ($L=100$ for AKLT, TB, and Heisenberg; $L=99$ for ULS)~\cite{liu2012}.}\label{tab:EgBBQ}
\end{table}

\begin{figure}[tpb]
	\includegraphics[width=8.4cm]{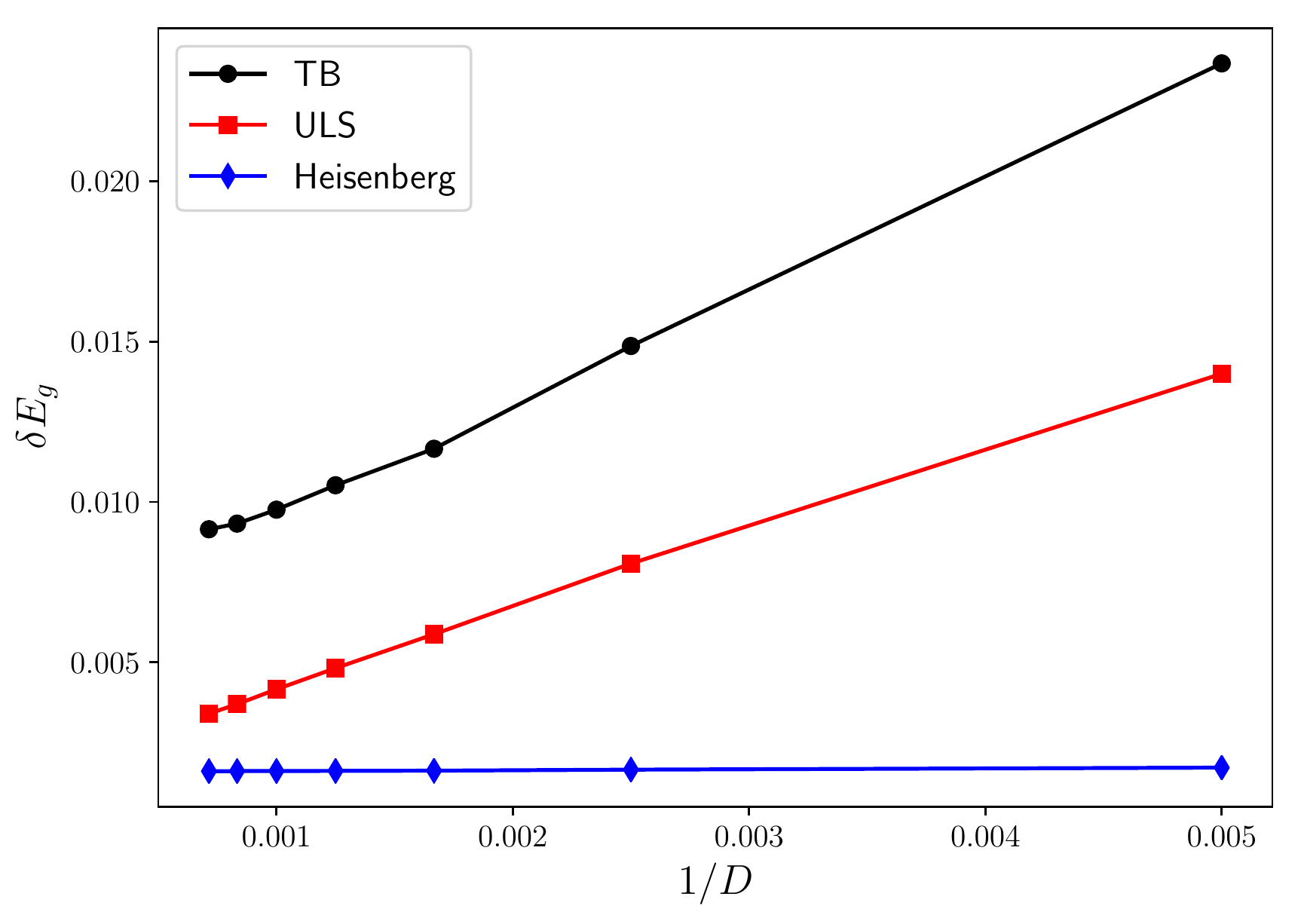}
	\caption{The energy deviations $\delta{}E_{g}$ defined in Eq.~\eqref{eq:edevia} versus $1/D$. The lattice size is $L=60$.}\label{fig:deviation}
\end{figure}

{\em  Energy deviation.---} The deviation of the ground-state energy measures the difference between  $E_{g}$ calculated by the basis-optimized MPO-MPS method and the known precise value $E_{g}^{*}$ given by exact solution (for AKLT~\cite{aklt}, TB~\cite{TB0,TB1} and ULS~\cite{bbq1d0,SU3ULSL,SU3ULSS}) or DMRG (for Heisenberg~\cite{S1dmrg}), which is defined as
\begin{equation}\label{eq:edevia}
\delta{}E_g(L,D)={E}_{g}(L,D) - E_{g}^{*}.
\end{equation}
Due to the lack of finite-size data, $E_g^{*}$ shown in Table~\ref{tab:errL60} is taken to be per-site energies in the thermodynamic limit ($L \rightarrow \infty$). The bond dimension dependence of $\delta{}{E}_{g}(L,D)$ reflects how fast ${E}_{g}(L,D)$ approaches the precise value ${E}_{g}(L,D=\infty)$ with increasing $D$, and is plotted in Fig.~\ref{fig:deviation}. It can be seen from Fig.~\ref{fig:deviation} that ${E}_{g}(L,D)-{E}_{g}(L,D=\infty)\propto 1/D$ for two gapless models, say, TB and ULS.
{\color{black}
For the ULS point, the parton wave function with $(\Delta/\chi,\lambda/\chi)=(0,1)$ is the exact ground state of the $SU(3)$ Haldane-Shastry model~\cite{SUN_HS}, which is known to be pretty close to the ground state of the ULS model.
However, the quality of the parton wave function is less clear for the TB point. As indicated in Fig.~\ref{fig:deviation}, the larger energy deviation for the TB point shows that there might still be room for improving the corresponding parton trial wave function.
}

{\em  Entanglement entropy. ---} The calculation of the von Neumann entanglement entropy $S=-\text{Tr}\rho_{A}\log\rho_{A}$ is straightforward in the MPO-MPS method. By contrast, it is unavailable in the VMC method although the R\'{e}nyi entropy $S^{(n)}=\text{Tr}\rho^{n}_{A}$ can be computed for $n=2$.

The AKLT state is an explicit spin wave function that realizes the Haldane phase for integer spins, whose ground state can be expressed as an MPS state~\cite{QSLRMP}. On the other hand, it was found that the spin-1 AKLT state can be exactly written as a Gutzwiller projected paired wave function $|\Psi_{\text{BBQ}}(\chi,\Delta,\lambda)\rangle$ by choosing $\chi=\Delta=1$ and $\lambda=0$~\cite{liu2012}. Our MPO-MPS calculation results in $E_{g}=-2/3$ and $S_{c}=2\log2$ exactly (in the sense of machine precision), where $S_{c}$ is the half-chain entanglement entropy defined in Eq.~\eqref{eq:Sc}. These results coincide with the exact solution.

For the two critical points, TB, and ULS, the von Neumann entanglement entropy $S$ is exploited to study the criticality as well. For the ground state of a 1D quantum critical chain with periodic boundary condition, the entanglement entropy between one block with $j$ spins and the other block with $L-j$ spins is known to scale as~\cite{cft0,cft1,cft2}
\begin{equation}\label{eq:cc}
S(j)=\frac{c}{3}\log\left(\frac{L}{\pi}\sin\frac{\pi{}j}{L}\right) + c_2,
\end{equation}
where $c$ is the central charge of the conformal field theory and $c_2$ is a nonuniversal constant.

\begin{figure}
	\includegraphics[width=8.4cm]{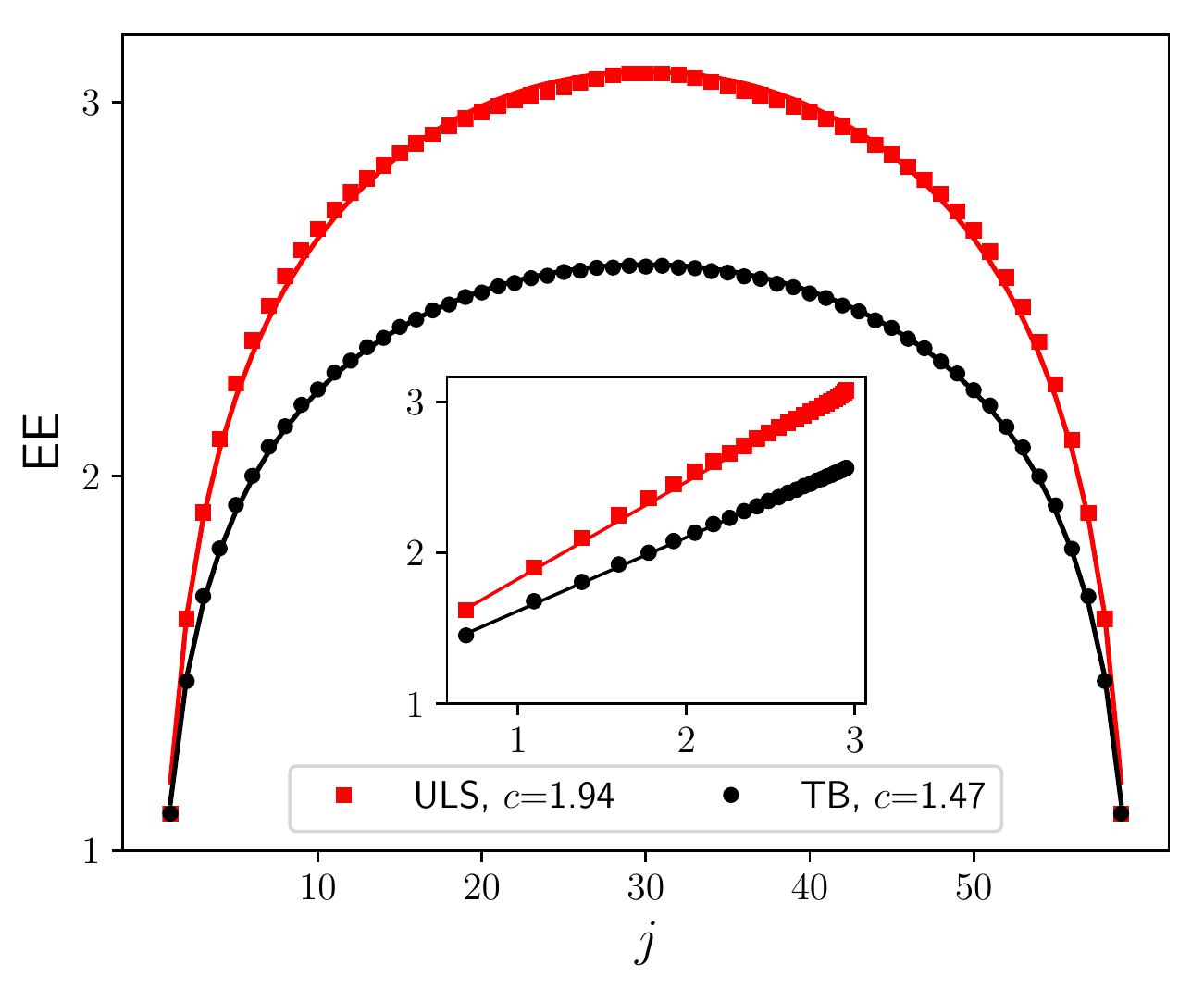}
	\caption{Block entanglement entropy $S(j)$ in an $L=60$ chain. The bond dimension in the MPO-MPS process is $D=1400$. For two gapless points, TB and ULS, central charges are fitted by Eq.~\eqref{eq:cc}.}\label{fig:ee}
\end{figure}

The entanglement entropy $S(j)$ has been calculated for two gapped states, AKLT and Heisenberg, and two gapless states, TB, and ULS, respectively. The results are plotted in Fig.~\ref{fig:ee}. The TB point is exactly solvable by Bethe ansatz~\cite{TB0,TB1}. By fitting $S(j)$ to Eq.~\eqref{eq:cc}, we find that the central charge extracted from the MPS is $c=1.47$, which agrees well with the theoretical value $c=3/2$ predicted by the $SU(2)_2$  Wess-Zumino-Novikov-Witten (WZNW) field theory~\cite{su2_20,su2_21}. The ULS point can be exactly solved by Bethe ansatz as well~\cite{SU3ULSL,SU3ULSS}, of which the low-energy excitations are effectively described by the $SU(3)_1$ WZNW field theory~\cite{suN_NLSM1,suN_NLSM2} with central charge $c=2$~\cite{suN_12}. Our MPO-MPS calculation from the Gutzwiller projected wave function leads to $c=1.94$, which is also in good agreement with the $SU(3)_1$ WZNW field theory.

\begin{figure*}[tpb]
	\includegraphics[width=12.8cm]{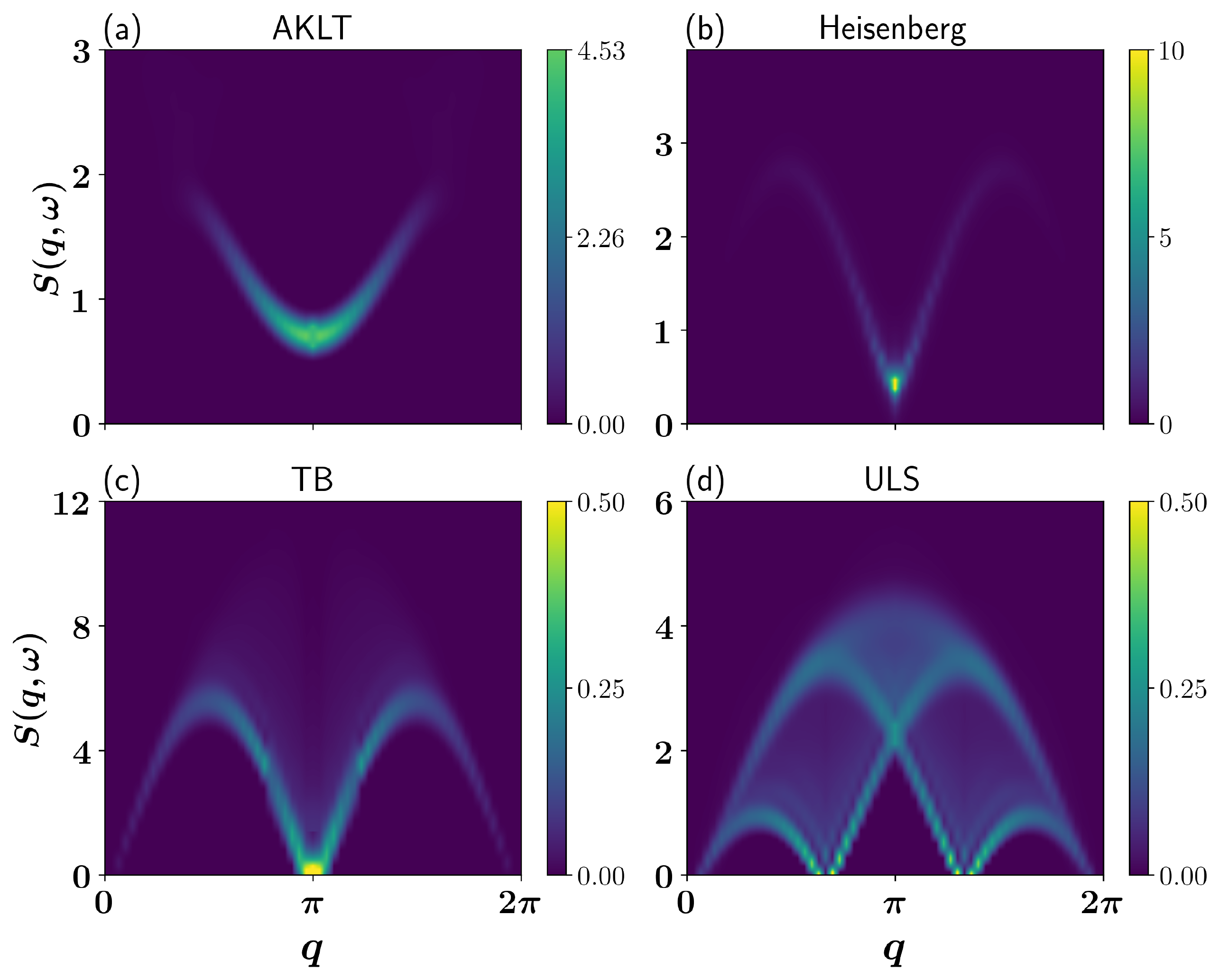}
	\caption{The spin spectral functions $S(\bm{q},\omega)$ for the spin-1 models. Here we set $J=1$, lattice size $L=60$, bond dimension $D=400$, and the number of Chebyshev moments $N_c=300$ ($N_c=150$ for TB point). (a) AKLT point, the Haldane gap at $\bm{q}=\pi$ is $\Delta_{\mathrm{gap}}=0.70$.{black} Heisenberg point, the Haldane gap at $\bm{q}=\pi$ reads $\Delta_{\mathrm{gap}}=0.39$. (c) TB point, $S(\bm{q},\omega)$ is gapless at $\bm{q}=\pi$ and there exists spinon continuum in the spectra. (d) ULS point, $S(\bm{q},\omega)$ is gapless at $\bm{q}=\pm{}2\pi/3$ and shows three thresholds in the spinon continuum.  }\label{fig:Sqw}
\end{figure*}

{\em  Spin spectral function. ---} The spin spectral function $S(\bm{q},\omega)$ has been calculated at AKLT, Heisenberg, TB, and ULS points by using the Chebyshev MPS method~\cite{chebyshevmps}. The numerical results are plotted in Fig.~\ref{fig:Sqw}, which catches all the expected features for these models.
(i) For AKLT and Heisenberg points, the Haldane gap is clearly visible at $\bm{q}=\pi$, and takes the value $\Delta_{\mathrm{gap}}=0.70$ at the AKLT point and $\Delta_{\mathrm{gap}}=0.39$ at the Heisenberg point.
(ii) For the TB point, the spin spectra are gapless at $\bm{q}=\pi$ and there exists spinon continuum in the spectra, which is exactly what is expected by the Bethe ansatz solution~\cite{tbspectrum}.
(iii) For the ULS point, $S(\bm{q},\omega)$ is gapless at $\bm{q}=\pm{}2\pi/3$ and exhibits three thresholds in the spinon continuum as shown in Fig.~\ref{fig:Sqw}~(d), which is in excellent agreement with the Bethe ansatz solution~\cite{SU3ULSS,ulsspectrum}.

\section{Summary and Discussions}\label{sec:summary}

In summary, we have generalized the basis-optimized MPO-MPS method to study Gutzwiller projected states of paired fermions. The key idea is that a BCS-type state can be obtained by filling up all the Bogoliubov quasiholes. Exploiting the maximally localized Wannier orbitals for the Bogoliubov quasiparticle/quasihole states, we are able to minimize the truncation error of MPS to improve the precision of this method dramatically.

As a benchmark, we have examined the 1D transverse field XY model, which is exactly solvable with the help of the Jordan-Wigner transformation. We found that the MPO-MPS method with maximally localized Wannier orbitals, together with the ``left-meet-right" scheme, gives rise to very accurate ground states. The precision for the per-site ground-state energy is always less than $10^{-10}$ in the whole parameter region inspected.

Then we used the Gutzwiller projected wave functions proposed in Ref.~[\onlinecite{liu2012}] to study $SO(3)$-symmetric spin-1 chains. We carefully examined the truncation error of MPS in the basis-optimized MPO-MPS process. The ground state and its variance have been evaluated with high precision. The von Neumann entanglement entropy has been calculated in a straightforward way. At the two critical points, TB and ULS, the central charges have been obtained by fitting the entanglement entropy: $c=1.47$ at the TB point and $c=1.97$ at the ULS point, which are in good agreement with $SU(2)_{2}$ and $SU(3)_1$ WZNW field theories, respectively. The spin spectral function $S(\bm{q},\omega)$ has been calculated at two gapped points, AKLT and Heisenberg, and two gapless points, TB and ULS. The Haldane gap was estimated to be $\Delta_{\mathrm{gap}}=0.39$ at the Heisenberg point and $\Delta_{\mathrm{gap}}=0.70$ at the AKLT point. The gapless feature has been found at TB and ULS points at $\bm{q}=\pi$ (TB) and $\bm{q}=\pm 2\pi/3$ (ULS), respectively.

{\color{black} We note that the form of MPO in Eq.~\eqref{eq:MPO_dm} is independent of spatial dimensionality, which indicates that one can apply the MPO-MPS method to quasi-1D systems in the same spirit as DMRG. If the parton wave function already captures the essential physics of the target Hamiltonian, the MPS obtained from such a Gutzwiller projected state would serve as a good initial state for DMRG simulations. Comparing with a random search from the beginning, this might save considerable computational costs.
}

{\color{black}  It is also worth mentioning again that some short-range RVB states have exact PEPS representations~\cite{Verstraete06a,Schuch12,Wang13,Poiblanc13} and can be written as projected BCS states explicitly~\cite{yang12,wildeboer12}. Thus, it is natural to expect that our MPO-MPS methods would work for these short-range RVB states. Moreover, the advantage of our MPO-MPS method is that we may deal with many long-range RVB states in the framework of projected BCS states, whose PEPS representation is unknown so far.

}

With minor modifications, the MPO-MPS method can also be utilized to represent partially Gutzwiller projected states, which allows us to study doped Mott insulators and strongly correlated metals.

Finally, there are some remaining issues for future investigations: (1) Is the tensor network representation for Gutzwiller projected states still efficient in dimensions larger than one? (2) It is natural to combine the MPO-MPS method with other techniques, such as automatic differentiation~\cite{AutoDiff}, to optimize the variational parameters in Gutzwiller projected wave functions. (3) Can a Gutzwiller projected state of bosons be calculated by the MPO-MPS method efficiently? The study along these lines are in progress.

\section*{Acknowledgment}
We thank Lei Wang, Zheng-Xin Liu, Xi Dai, and Ying-Hai Wu for helpful discussions.
This work is supported in part by National Natural Science Foundation of China (No. 11774306), National Key Research and Development Program of China (No.2016YFA0300202), the Strategic Priority Research Program of Chinese Academy of Sciences (No. XDB28000000) and the DFG through project A06 of SFB 1143 (project-id 247310070).

\appendix

\begin{appendices}

\section{Chebyshev MPS approach to spin spectral function}\label{app:chbmps}
This appendix briefly reviews the Chebyshev MPS method~\cite{chebyshevmps} for calculating the zero-temperature spectral functions.
Here we use the spin spectral function $S(\bm{q},\omega)$ as an example, which has the form of Eq.~\eqref{eq:Sqw}.

The key intuition of the Chebyshev MPS method is representing the spectral function via the Chebyshev expansion, which is widely used for function expansion.
The Chebyshev polynomials constitute an orthogonal basis with a weight function $(\pi\sqrt{1-x^{2}})^{-1}$ on the interval $[-1,1]$. So at first the frequency $\omega$ and Hamiltonian $\hat{H}$ should be linearly rescaled and shifted~\cite{rmp_Fehske} to map the interval $[0,W_*]$, such that the spectral function has nonzero weight in the interval $[-W^\prime,W^\prime]$, where $W^\prime<1$ is a positive real number.
The rescaled dimensionless Hamiltonian and frequency are marked by primes, such as
\begin{equation}
\begin{split}
\omega^{\prime}&=\frac{2W^\prime}{W_*}\omega{}-W^\prime,\\
\hat{H}^\prime&=\frac{2W^\prime}{W_*}(\hat{H}-E_0)- W^\prime,
\end{split}
\end{equation}
where the ground-state energy of $\hat{H}^\prime$ is $W^\prime$.
Then, $S(\bm{q},\omega)$ can be rewritten as
\begin{equation}\label{eq:Sqwprime}
S(\bm{q},\omega)=\frac{2W^\prime}{W_*}\sum_{a}\langle{}0|S ^{a}(\bm{q})\delta(\omega^\prime-\hat{H}^\prime)S^{a}(-\bm{q})|0\rangle.
\end{equation}
The $\delta(\omega^\prime-\hat{H}^\prime)$ in Eq.~\eqref{eq:Sqwprime} is expanded by Chebyshev polynomials as
\begin{equation}\label{eq:Chebyexpansion}
\delta(\omega^{\prime}-\hat{H}^\prime)\approxeq\frac{1}{\pi\sqrt{1-\omega^{\prime{}2}}}\left[g_0+2\sum_{n=1}^{N_c-1}g_nT_n(\hat{H}^\prime)T_n(\omega^{\prime})\right],
\end{equation}
where $T_n(x)\equiv{}\cos[n\arccos(x)]$ are the Chebyshev polynomials of the first kind and $g_n$ damping factors~\cite{chebyshevmps,rmp_Fehske}.
It is worth noting that Eq.~\eqref{eq:Chebyexpansion} is an approximation for $\delta$ function since only the first $N_c$ Chebyshev moments are retained, and such a truncation introduces the Gibbs oscillations\cite{rmp_Fehske} of period $1/N_c$. So here the Jackson damping
\begin{equation}
g^{J}_n=\frac{(N_c-n+1)\cos\frac{\pi{}n}{N_c+1}+\cot\frac{\pi{}}{N_c+1}\sin\frac{\pi{}n}{N_c+1}}{N_c+1}
\end{equation}
is adapted to smooth the oscillations.

Conclusively, the spin spectral function $S(\bm{q},\omega)$ can be represented approximately as
\begin{equation}
S(\bm{q},\omega)\approxeq\frac{2W^\prime/{W_*}}{\pi\sqrt{1-\omega^{\prime{}2}}}\left[g_0\nu_0+2\sum_{n=1}^{N_c-1}g_n\nu_nT_n(\omega^{\prime})\right],
\end{equation}
where
{\color{black}
\begin{equation}
\begin{split}
\nu_n&=\sum_{a}\langle{}0|S^{a}(\bm{q})T_n(\hat{H}^\prime)S^a(-\bm{q})|0\rangle\\
&=\sum_{a}\langle{}0|S^{a}(\bm{q})|t_n^{a}(\bm{q})\rangle
\end{split}
\end{equation}
is so-called {\em Chebyshev moments} which are obtained from the {\em Chebyshev vectors}
$$|t_n^{a}(\bm{q})\rangle=T_n(\hat{H}^\prime)S^a(-\bm{q})|0\rangle.$$
Note that Chebyshev polynomials, $T_n(x)$, have the recurrent relations of
\begin{equation}\label{eq:rrelation}
T_{n+1}(x)=2xT_n(x)-T_{n-1}(x).
\end{equation}
Thus $|t_n^{a}(\bm{q})\rangle$ can be calculated recursively and efficiently by using Eq.~\eqref{eq:rrelation} as
\begin{equation}\label{eq:t_n}
|t_n^{a}(\bm{q})\rangle=2\hat{H}^\prime|t_{n-1}^{a}(\bm{q})\rangle-|t_{n-2}^{a}(\bm{q})\rangle,
\end{equation}
where
$$|t_0^{a}(\bm{q})\rangle=S^a(-\bm{q})|0\rangle,\qquad|t_1^{a}(\bm{q})\rangle=\hat{H}^\prime|t_0^{a}(\bm{q})\rangle.$$
The first-order Chebyshev vector $|t_1^{a}(\bm{q})\rangle$ and the higher orders in Eq.~\eqref{eq:t_n} can be implemented by the standard compression procedure~\cite{schollwock11}, i.e., by variationally minimizing
\begin{equation*}
\|\,|t_1^{a}(\bm{q})\rangle-\hat{H}^\prime|t_{0}^{a}(\bm{q})\rangle\,\|_2^2
\end{equation*}
and
\begin{equation*}
\|\,|t_n^{a}(\bm{q})\rangle-2\hat{H}^\prime|t_{n-1}^{a}(\bm{q})\rangle+|t_{n-2}^{a}(\bm{q})\rangle\,\|_2^2.
\end{equation*}
Here the 2-norm distance is defined as
\begin{equation*}
\|\,|\psi\rangle-|\tilde{\psi}\rangle\,\|_2^2=\langle\psi|\psi\rangle+\langle\tilde{\psi}|\tilde{\psi}\rangle-\langle\psi|\tilde{\psi}\rangle+\langle\tilde{\psi}|\psi\rangle.
\end{equation*}
}
\end{appendices}
\bibliography{mpomps}

\end{document}